# Origin of the anomalous low temperature upturn in resistivity in the electron-doped cuprates.

Y. Dagan [1], A. Biswas [2], M. C. Barr [1], W. M. Fisher [1], and R. L. Greene [1].

[1] Center for Superconductivity Research, Department of Physics, University of Maryland, College Park, Maryland 20742, USA.

[2] Department of Physics, University of Florida, Gainesville, Florida 32611.

The temperature, doping and field dependences of the magnetoresistance (MR) in $Pr_{2-x}Ce_xCuO_{4-\delta}$ films are reported. We distinguish between orbital MR, found when the magnetic field is applied perpendicular to the ab planes, and the nearly isotropic spin MR. The latter, the major MR effect in the superconducting samples, appears in the region of the doping-temperature phase diagram where $d\rho/dT<0$, or an upturn in the resistivity appears. We conclude that the upturn originates from spin scattering processes.

Fermi liquid theory generally describes the normal state of conventional superconductors. In the high-$T_c$ cuprates, both the normal and the superconducting states depend on the carrier concentration in the $CuO_2$ planes (doping). In hole doped (p-doped) cuprates, the overdoped region is believed to be metallic (Fermi liquid-like), whereas in the underdoped region, at low temperatures the resistivity increases with decreasing temperatures and may even be logarithmically diverging at T=0 [1]. A similar behavior with decreasing doping is found in the electron-doped (n-doped) cuprates [2]. For optimally doped and underdoped samples when T is decreased from high temperature the resistivity reaches a minimum at $T_{min}$ and then begins to increase. $T_{min}$ increases with decreasing doping. The origin of this anomalous upturn in resistivity has not been determined and this is the main subject of this paper.

The normal state of the n-doped cuprates is also characterized by negative magnetoresistance (n-MR) at low temperatures. Fournier *et al.* [3] interpreted the upturn in resistivity, as well as the n-MR, as a result of two dimensional (2D) weak localization by disorder. In contrast, Sekitani *et al.* [4] suggested that the resistivity upturn and the n-MR are due to scattering off $Cu^{2+}$ Kondo impurities induced by residual apical oxygen. In the p-doped cuprates, for example, negative MR was found in underdoped $La_{2-x}Sr_xCuO_4$ (LSCO) [5]. Other work was mainly focused on in-plain MR anisotropy in lightly doped LSCO [6], $Pr_{1.3-x}La_{0.7}Ce_xCuO_4$ [7] or non superconducting $Pr_{2-x}Ce_xCuO_{4-\delta}$ (PCCO) x=0.15 [8].

Recently, we found that the normal state Hall coefficient of PCCO at 300mK exhibits a dramatic change at x=0.165±0.005. [9] This singular behavior was accompanied by significant changes in the temperature dependence of the resistivity below 20K. These changes in the resistivity and Hall coefficient were suggested as strong evidence for a quantum critical point (QCP) at $x_c$=0.165±0.005. We also found that below $x_c$ the upturn in resistivity appears at low temperatures. The broad antiferromagnetic (AFM) region from x=0 to just above x=0.15 found in the phase diagram of the n-doped cuprates [10, 11, 12] suggests that the QCP found in ref. [9] can be associated with the disappearance of the AFM phase as the doping is increased at T=0.

In this work we report a new effect; an almost isotropic, spin related MR in PCCO. While an orbital MR exists in the whole doping range and at a far different temperature scale than that of the resistivity upturn, the spin component exists only below x=0.16 and it has the same temperature scale as the upturn. *We therefore conclude that the resistivity upturn is due to spin scattering*. It may be related to AFM, which is found to persist well into the superconducting dome.[11]

The samples are *c*-axis oriented PCCO thin films: x=0.11, 0.12, 0.13, 0.15, 0.17, 0.18 and 0.19, whose preparation and characterization procedures were described elsewhere.[9] Measurements in the National High Magnetic Field Lab (NHMFL) were taken in a 32.4 T magnet and at temperatures ranging from 1.5 K to 20 K. Other measurements were taken using a Quantum Design PPMS 14T magnet. The field was aligned parallel to the ab planes (H∥ab) with an accuracy better than 0.25°. To exclude eddy current heating effects we ensured that the data was reproducible, symmetric for positive and negative magnetic fields and independent of the sweeping rate. We measure the ab-plane resistivity with a standard 4-probe technique.

In Fig. 1 we show the field dependence of the resistivity for two doping levels, x=0.15 and 0.16, with field applied perpendicular to the ab planes (H⊥ab). Above a certain field, necessary to mute superconductivity, both samples exhibit n-MR at low temperatures. The amplitude of the n-MR decreases as the temperature increases and eventually vanishes around 8-10K. The n-MR is found in the entire doping range studied (data not shown).

An immediate conclusion drawn from Fig.1, and from the existence of n-MR for H⊥ab for all x studied, is that there is no direct relation between the n-MR for H⊥ab and the upturn in resistivity. First, the upturn in resistivity appears only for $x \leq 0.16$ while the n-MR with H⊥ab appears for $0.11 \leq x \leq 0.19$. Second, the upturn in resistivity and the onset of the n-MR with H⊥ab occur at unrelated temperatures. For example, in x=0.15 the $T_{min} \approx 20K$, while the n-MR with H⊥ab vanishes around 8K. For the x=0.16 we find $T_{min} < 1K$, while the n-MR for H⊥ab is observed up to 8K. The n-MR for H⊥ab will be discussed in a separate publication. In this paper, we will concentrate on the negative longitudinal MR (LMR) due to the field component parallel to the *ab* planes. We show that the LMR is closely related to the resistivity upturn: it vanishes at the same doping level and approximately the same temperatures at which the upturn disappears.

The resistance versus field, H||ab, is shown in Fig. 2a for the nonsuperconducting ($T_c < 0.35K$) x=0.11 film at 1.8K. We now follow this LMR as a function of temperature and doping. In Fig. 3a we show the LMR as a function of field at various temperatures for the x=0.13 sample. The LMR is negative at low temperatures and changes to positive above 80K. The same behavior is observed in the x=0.11 sample but with a crossover from negative to positive LMR around 100K. In Fig. 3b we show the resistivity for x=0.13 as a function of temperature; note that $T_{min} \approx 60K$. However, it is difficult to say exactly where the upturn begins. To give a better estimate for this temperature we plot the effective resistivity exponent, $\beta = \frac{d \ln(\rho - \rho_0)}{d \ln T}$ as a function of temperature (circles, right hand scale Fig. 3b). $\rho_0$ is calculated from fitting the resistivity in the range 120K< T < 250K to the form $\rho = \rho_0 + AT^2$. Around 90K β starts falling rapidly from its value at high temperatures, β=2, to 0 at the minimum. This temperature is very close to the temperature at which LMR crosses over from positive to negative. The same result was also found in the x=0.11 film where the upturn begins around 100K, approximately where the LMR appears. For higher dopings it is difficult to make such a measurement since the upturn begins close to (x=0.15) or below $T_c$ (x=16), and a parallel field of 14T is not sufficient to mute the affects of the superconductivity.

We have demonstrated that the upturn in resistivity and the LMR appear at the same characteristic temperature. We now show that they also have the same characteristic doping. Since the upper critical field for PCCO for H||ab is greater than 32T it is impossible to directly measure the LMR and study its doping dependence at low temperatures. However, we can carry out another procedure that gives similar information. First, we measured the resistance as a function of H⊥ab up to 32.4T at 1.5K. We then rotate the film in a field of 32.4T with θ the angle between the field and the ab-planes. θ=0 corresponds to field applied parallel to the current; θ=90° is field perpendicular to the film. If only $H_\perp$ affects the MR, then the resistance at a field $H_0$ applied perpendicular to the film should be the same as when the maximal available field of 32.4T is applied at an angle $\theta_0 = \arcsin(H_0/32.4T)$. We therefore transform the field sweep into an effective angle, θ, using $\theta = \arcsin(H/32.4T)$.

The H⊥ab sweep and the rotation in field measurements are compared in Fig.4a. In the high doping regime (x=0.16) the results of the field sweep (green squares) and those from the rotation in field (black circles) almost overlap. This means that there is only a perpendicular MR component i.e. orbital MR. Similar results were obtained for the x=0.17 and x=0.19 samples (not shown). A completely different behavior is found in the x=0.15, x=0.13, x=0.12, and x=0.11 samples. There we can identify a strong parallel magnetic field effect, observed as the difference between the field sweep and the rotation in field. In fact, the MR is almost isotropic in the x=0.13, x=0.15 samples (black circles). For example in x=0.15 the resistivity changes by about 8% in the H⊥ab sweep while it changes by only 1% in the rotation. Surprisingly, this suggests that most of the n-MR seen when H⊥ab is not due to orbital or 2D effects such as weak localization [3] or to a field-tuned 2D superconductor-to-insulator transition [13].

Based on the above results we assume the existence of an H⊥ab effect (orbital) and an isotropic effect (spin). We use the x=0.11 sample, where absence of superconductivity allows us to measure the resistance as a function of parallel (Fig.2a) and perpendicular fields from 0 to 14T and the full rotation in a 14T field to check this assumption (see Fig. 2b). The orbital effect is simply $\Delta R_{orb}(H)=\Delta R(H_{\perp})-\Delta R(H_{||})$, since the parallel field picks up only the isotropic component. From $\Delta R_{orb}(H)$, one can calculate $\Delta R_{orb}(\theta, H=14T)$ using the previous procedure for converting field sweeps to θ sweeps (red dotted line Fig. 2b.), and obtain $\Delta R_{spin}(\theta, H=14T)=\Delta R_{measured}(\theta, H=14T)-\Delta R_{orb}(\theta, H=14T)$. $\Delta R_{measured}(\theta, H=14T)$ (dashed green line) and the calculated $\Delta R_{spin}$ at 14T (black solid line) are plotted in Fig. 2b as a function of angle. We note that $\Delta R_{spin}$ is almost isotropic, consistent with a spin scattering mechanism. This is a verification of our assumption.

Finally, in fig.4b we plot the difference between the resistance at 16.2T (H⊥ab), (effective angle θ=Arcsin(16.2/32.4)=30°) and the resistance at 32.4T applied at 30°, normalized with the resistance at 32.4T (H⊥ab) versus doping (all at T=1.5K). This is a measurement of the isotropic spin effect of the additional 16.2T. The spin effect vanishes dramatically at x=0.16 at this temperature. This is consistent with the evidence for a quantum phase transition at x=0.165±0.005 that we have previously reported [9]. In the x=0.16 sample, the upturn can be seen only below 1K, hence, the absence of a LMR component in this sample at 1.5K is not surprising and it is still consistent with a phase transition at T=0 and $x_c$=0.165±0.005. The sudden increase of Δρ/ρ at x=0.15 might be due to enhanced scattering near the hotspots observed in ARPES [14].

We have shown here a correlation between the upturn and the isotropic, spin related MR, and therefore, we suggest that the upturn is a result of a spin effect. A straight forward explanation could be a partial gapping of the Fermi surface due to AFM correlations as seen in angular resolved photoemission spectroscopy [14] and in optics [15]. However, the Nèel temperatures reported for x=0.10, 0.13 are 150K, 120K respectively. These temperatures are higher than the upturn/LMR temperatures that we find here. On the other hand, Woods *et al.* [16] irradiated an optimally-doped NCCO film and found an increase of $T_{min}$ with increasing disorder. Therefore, we speculate that the upturn and the spin MR are related to impurities. In the non-magnetic phase the samples are less susceptible to impurities and therefore exhibit no LMR and no upturn. However, on the AFM side of the phase diagram, magnetic droplets can nucleate around impurities and cause an increase in resistance. Scattering off these droplets is suppressed by the magnetic field and hence the negative LMR. More quantitative calculations will be needed to confirm this suggestion.

In summary, doping, temperature, and field dependences of the ab plane magnetoresistance were measured in $Pr_{2-x}Ce_xCuO_{4-\delta}$ epitaxial films. An orbital magnetoresistance component exists throughout the whole doping range measured, $0.11 \leq x \leq 0.19$. Its temperature dependence is very different from that of the anomalous resistivity upturn. By contrast, the longitudinal magnetoresistance, measured with field applied parallel to the ab planes, is observed in the same doping and temperature ranges as the resistivity upturn. We show that this component is better described by isotropic spin scattering. We conclude that the anomalous upturn in resistivity is due to spin scattering. We speculate that this scattering is due to magnetic droplets formed around impurities on the AFM side of the phase diagram.

We are indebted to A. J. Millis for useful discussions and for suggesting the magnetic droplets scenario, to A. V. Chubukov, R. A. Webb, J. S. Higgins for useful discussions, and to R. Beck, T. Dhakal and S. Hannas for their help at the NHMFL. NSF grant DMR 03-52735 supported this work. Work carried out at the NHMFL is supported by an NSF cooperative agreement DMR-00-84173 and by the State of Florida.

Figure Captions.

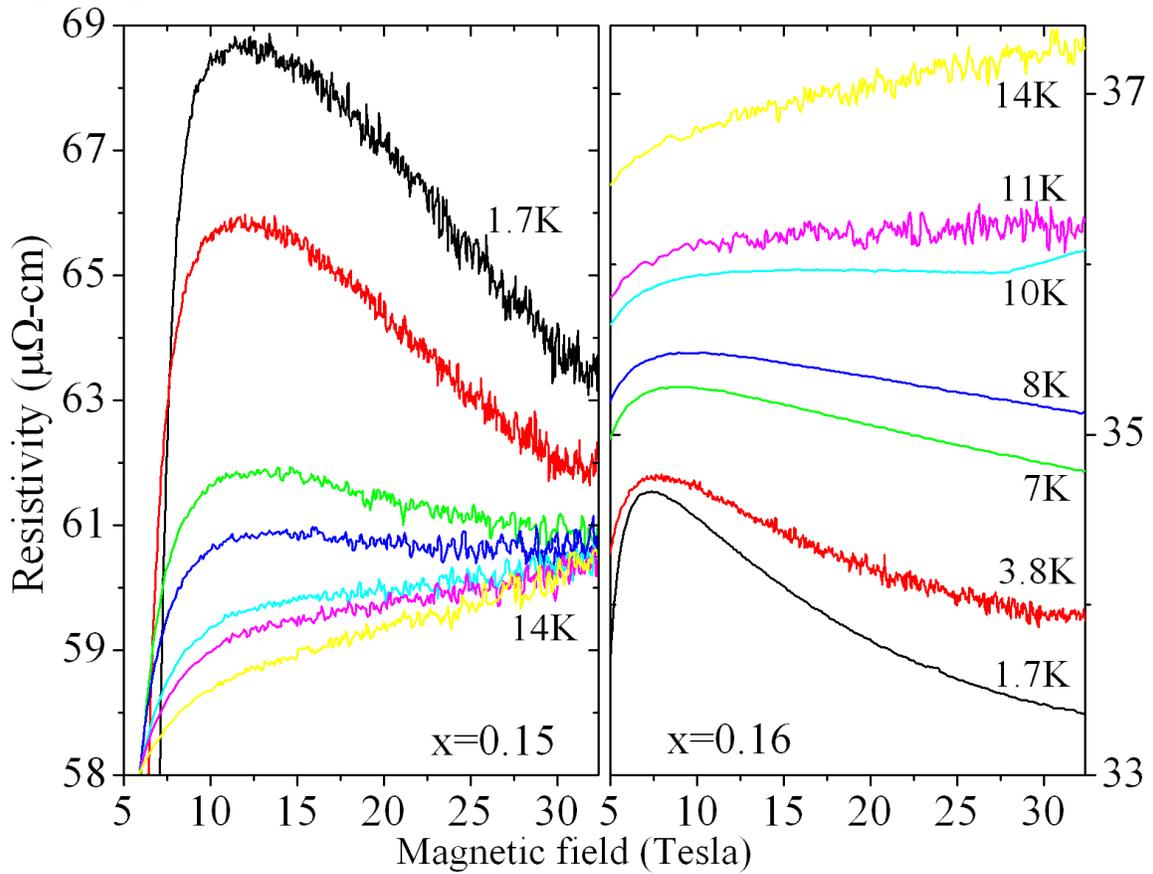

Figure 1.
The ab-plane resistivity of $Pr_{2-x}Ce_xCuO_4$ films vs. magnetic field applied perpendicular to the ab plane (H⊥ab).

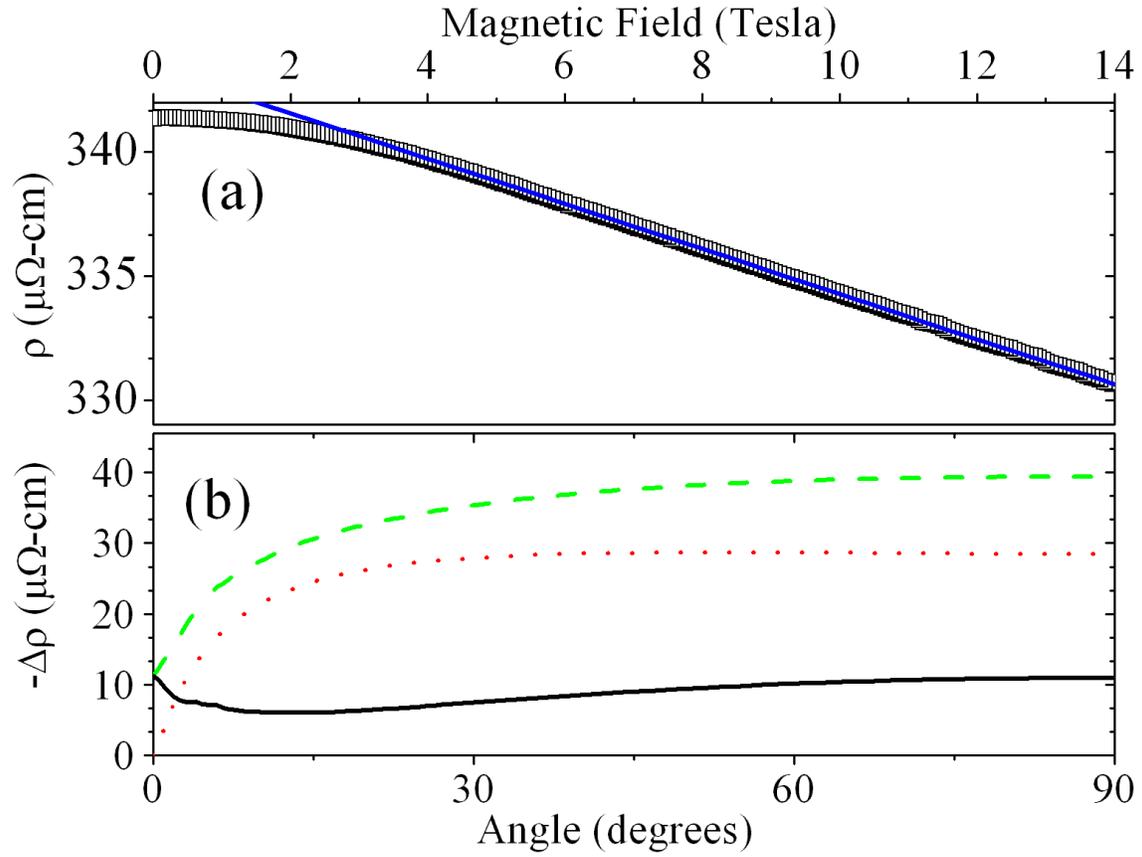

x=0.11 sample at T=1.8K. (a) The resistivity versus field at $\theta=0$ (H∥ab). Solid line is a linear fit to the high fields region. (b) Calculation of the angular dependence of a spin effect; see text for details.

Figure 3.

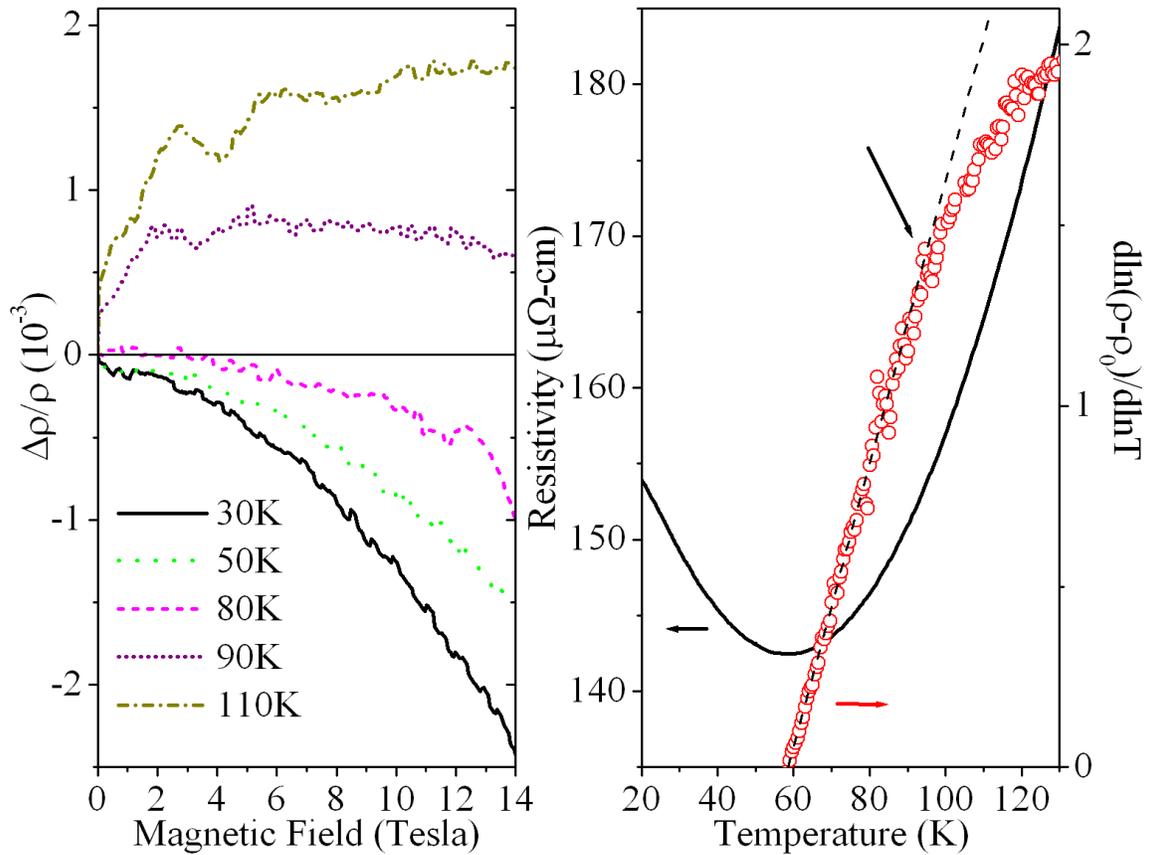

(a) The temperature dependence of the LMR (H||ab) for x=0.13 sample at various temperatures. Note that the MR turns positive above 80K. (b) The resistivity as a function of temperature for the same sample (solid line, left hand scale). The effective exponent dln($\rho-\rho_0$)/dlnT (circles, right hand scale) changes rapidly around 90K, i.e., where the LMR changes sign. The dashed line is a guide to the eye.

Figure 4.

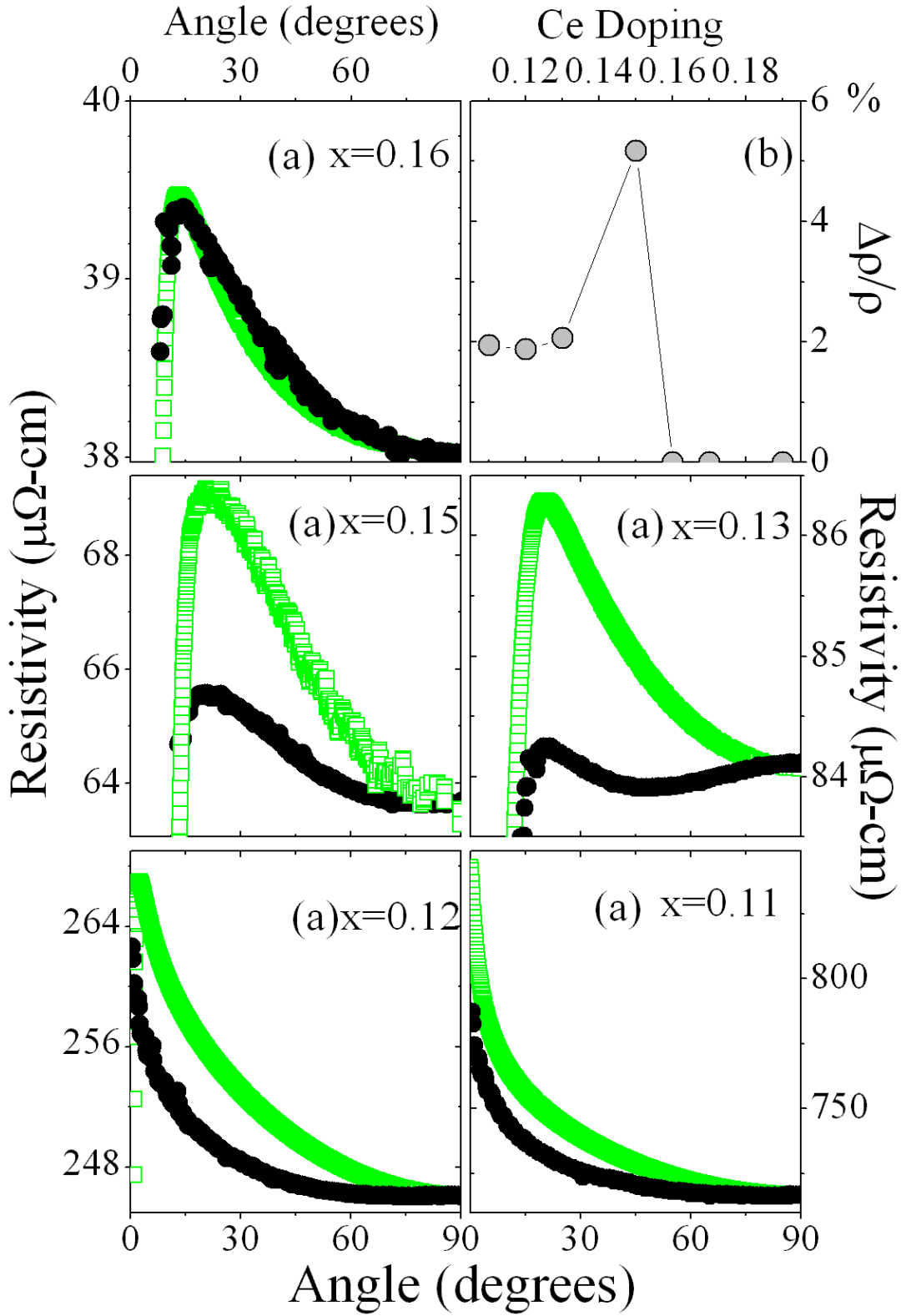

(a) Comparison between field sweeps H⊥ab (black circles) and rotation in a magnetic

field of 32.4T (green squares) for various dopings at 1.5K. The difference between the two measurements is due to the spin effect, which vanishes for x≥0.16 (see text).
(b) The difference in resistivity between the field sweep and the rotation in field at 30° taken from fig.4(a) and normalized with the resistivity at 32.4T (H⊥ab).